# Persuasive Technologies for Sustainable Urban Mobility

Evangelia Anagnostopoulou[1], Efthimios Bothos[1], Babis Magoutas[1], Johann Schrammel[2], Gregoris Mentzas[1]

[1]ICCS- Institute of Communication and Computer Systems, NTUA- National Technical University of Athens, Greece
{eanagn, mpthim, elbabmag, gmentzas}@mail.ntua.gr
[2]AIT – Austrian Institute of Technology, Vienna, Austria
Johann.Schrammel@ait.ac.at

**Abstract.** In recent years, the persuasive interventions for inducing sustainable urban mobility behaviours has become a very active research field. This review paper systematically analyses existing approaches and prototype systems and describes and classifies the persuasive strategies used for changing behaviour in the domain of transport. It also studies the results and recommendations derived from pilot studies, and as a result of this analysis highlights the need for personalizing and tailoring persuasive technology to various user characteristics. We also discuss the possible role of context-aware persuasive systems for increasing the number of sustainable choices. Finally, recommendations for future investigations on scholarly persuasive systems are proposed.

## 1 Introduction

Transport systems have significant impacts on the environment, accounting for between 20% and 25% of the world energy consumption and carbon dioxide emissions [1]. Current transportation practices are not sustainable, as recent reports show that the greenhouse gas (GHG) emissions from transport are increasing at a faster rate than any other energy using sector, especially in urban environments [2]. Modern cities suffer from over utilisation of land resources, increased urbanisation and mobility solutions that are highly dependent on private vehicles. This has produced highly congested urban environments and conditions detrimental to the quality of life of local inhabitants with adverse effects on public health and the environment.

In order to respond to these emerging unsustainable conditions, a broad range of strategies is required, including increasing vehicle efficiency, lowering the carbon content of fuels, and reducing vehicle miles of travel. Increasing travellers' aware-ness of the environmental impact of travel mode choices and changing the citizen's behaviour towards adopting transportation habits that rely more on the use of public transportation, bicycles and walking and less on private cars, can provide the means to reduce GHG emissions in the short term, and mitigate the effects on the environment. Other positive effects include less local air pollution and smog, as well as more healthy lifestyles, with increase exercise and less obesity [3].

In this context, persuasive technologies, tailored for and integrated in route planning applications, can affect urban travellers' decisions and guide them towards selecting routes that are environmentally friendly. Persuasive technology is broadly defined as technology that is designed to change attitudes or behaviours of the users through persuasion and social influence, but not through coercion [4]. Persuasive systems addressing behaviour change in the context of personal mobility in urban environments is an active area of research, and numerous example systems and implementations exist, aiming to motivate users towards making more eco-friendly choices.

Many approaches utilizing different strategies such as behaviour feedback, social comparison, goal-setting, gamification, personalized suggestions and challenges have been used so far, and new ones are continuously developed. Similar to the diversity of approaches also the implementation details (e.g. mobile trip planner app versus web-based systems) are very divergent. Last but not least the transportation context especially the available transportation infrastructure, possible trip alternatives and costs associated with the different transportation modes are defined by the targeted area and can vary substantially. Due to this kaleidoscope of influences and variables both researchers and practitioners may find it difficult to extract the main findings relevant for their own project or research interests.

In this paper we review persuasive systems implementations and related pilot studies, aiming to systematize available research results and provide a framework for understanding and interpreting approaches for persuasion in the context of personal mobility in urban environments. The paper is organized as follows. Section 2 presents the process followed for the review analysis. Section 3 describes the systems that have implemented while Section 4 provides an analysis of related pilot studies. Section 5 discusses the need of personalization, context awareness and proactivity in persuasive technologies and how it can be proceed. Finally, Section 6 concludes the paper with our final remarks and next steps.

## 2      Methodology

We used the methodology introduced in [5] that provides rigorous and well-defined guidelines for performing literature reviews. Firstly, we determined that there was no systematic review in the field of persuasion for sustainable mobility. The increasing number of papers on persuasion technologies for sustainable urban mobility is ample evidence that it had been an important issue in the last years. Identifying appropriate persuasive strategies and system designs to induce sustainable behaviours in transportation is needed for future studies in this field. Hence, there is a need to conduct a systematic review based on the results from the past studies that used different strategies and systems to persuade users to make more sustainable choices.

In order to find the relevant studies for the review we chose bibliographic databases that cover the majority of journals and conference papers published in the field of persuasive technologies and computer science in general. We selected the following bibliographic databases as relevant: ACM, Ebsco, Web of Science, Proquest, IEEE, ScienceDirect, SpringerLink, Emerald, Google Scholar, and Ingenta Journals. To retrieve the relevant papers, we carried out searches in these databases with the following combinations of keywords: "persuasive strategies", "sustainable mobility", "behavioural change". The total number of papers retrieved through the above method was 80.

We reviewed the papers through the abstract, introduction and conclusion. Among the retrieved papers, we selected papers that implement a persuasive application for sustainable urban mobility using one or more strategies and/or present results from pilot cases. Considering the above restrictions, we selected 12 papers for final review. In the following we analyse the systems (Section 3) and the pilot cases (Section 4) reported in these papers.

## 3      Persuasive Systems

Persuasive systems aim to change behaviour related to sustainability typically by raising individuals' awareness of their choices, behaviour patterns and the consequences of their activities. These systems incorporate one or more persuasive strategies in order to motivate change. During the review of related systems, we identified a set of five strategies that are commonly used. These are provided in Table 1, together with their definition, and will be used for the purposes of our analysis in the remainder of this paper.

**Table 1.** Identified persuasive strategies in urban mobility applications.

| Persuasive Strategy | Description [6] |
|---|---|
| Challenges & Goal Setting | Offering challenges and setting goals that incentivise the user to show an intended behaviour in a self-competitive context through a comparison of the present and a desirable future situation. |
| Self- monitoring | Applying computing technology to eliminate the tedium of tracking performance or status helps people to achieve predetermined goals or outcomes. |
| Personalized messages | Information provided by computing technology will be more persuasive if it is tailored to the individual's needs, interests, personality, usage context, or other factors relevant to the individual." |
| Social comparison | System users will have a greater motivation to perform the target behaviour if they can compare their performance with the performance of others. |
| Gamification & Rewards | To (virtually) reward target behaviours influences people to perform the target behaviour more frequently and effectively. |

Table 2 presents the persuasive systems we identified in the domain of mobility. One of the first attempts is the PerCues mobile app [7] which aims to persuade people to use public transportation instead of their car in order to reduce emissions. The approach is based on displaying personalized bus and pollution information, such as the departure time of the next bus and the decrease in emissions achieved by taking the bus instead of the car. Users can also see the impact of the actions of other users on the environmental pollution.

The UbiGreen app [8] encourages greener alternatives, including carpooling, public transport and pedestrian modalities by providing visual feedback in the form of adapting the background graphics of the smartphone when users reduce driving. UbiGreen makes use of sensors to semi-automatically infer transportation mode and monitor users' transportation behaviours.

**Table 2.** The persuasive systems for sustainable mobility we identified in the literature.

| Ref. | System | Type | Integrated Persuasive Strategies |
|---|---|---|---|
| [7] | PerCues | Mobile | Self- monitoring |
| [8] | UbiGreen | Mobile | Self- monitoring |
| [2] | TRIPZOOM | Mobile | Self- monitoring, Challenges & Goal Setting, Social comparison |
| [9] | QT | Mobile/web | Self- monitoring, Social comparison |
| [10] | MatkaHupi | Mobile | Self- monitoring, Challenges & Goal Setting |
| [12] | IPET | Mobile | Self- monitoring, Personalized messages |
| [11] | Peacox | Mobile/web | Self- monitoring, Challenges & Goal Setting, Social comparison, Personalized messages, Gamification & Rewards |
| [13] | SUPERHUB | Mobile | Self- monitoring, Challenges & Goal Setting, Social comparison, Personalized messages, Gamification & Rewards |
| [14] | Moves | Mobile | Self- monitoring, Social comparison |
| [15] | Viaggia Roveretgoto | Mobile | Gamification & Rewards |

TRIPZOOM [3] aims to optimize mobility by supporting users to gain insights on their mobility behaviour. The app tracks users' mobility patterns, allows them to zoom in on trip details, including costs, emissions and impact on health and provides rewards that incentivize users to save CO2 emissions. Moreover, it supports social comparisons by offering users functionalities to share achievements in social networks such as such as Facebook and Twitter.

Quantified Traveller (QT) [9] app provides a computational alternative to counsellors of travel feedback programs. It collects travel information and feeds it back to in the form of "augmented" travel diaries in order to encourage pro-environmental mobility. The presented information includes personalized carbon, exercise, time, and cost footprint, while the design also embodies social comparisons.

MatkaHupi [10] is a journey planning app that detects and records users' trips and transport modes and provides eco-feedback in the form of CO2 emissions visualiza-tions. Moreover, the app employs a gamification approach for persuasion in the form of challenges offered to users based on their observed behaviour. After a trip, the app checks for faster and/or with lower emissions alternatives which are presented to the user. S/he is then challenged to consider the proposed alternative in the future and reward with points if the challenge is accepted and achieved.

Peacox [11] influences urban travellers to consider the environmental friendliness of travel modes while planning a route. A choice architecture approach nudges users to shift to less polluting modes by filtering and structuring the alternative routes according to user preferences and contexts while emphasizing the environmentally friendly routes. Moreover the app embeds CO2 emissions visualizations as well as personal and collaborative challenges aiming to persuade users to reduce the emissions caused by their mobility choices.

The IPET platform [12] integrates functionalities for the provision of persuasive information and advice to mobile devices. More specifically, it tracks user activities, analyses them to detect the used trip mode and infers alternative and more sustainable routes which are communicated to the user using brief persuasive messages that combine text and images in different, including comics and real life sceneries.

SUPERHUB [13] is a mobile app which motivates users making more sustainable choices using a novel combination of goal-setting, self-monitoring, rewards and sharing features. It supports multi-modal journey planning, personalized recommendations, and behaviour change for environmentally sustainable travel. There are also many functionalities, such as event reporting, social media and transport data-feed scanning that aim to more self-contained, comprehensive and accurate user experience.

Moves [14] is an activity tracking app that provides data on the user's time and distance spent under each active mode, using a combination of accelerometer and location data to distinguish between motorized transportation cycling, walking, and running,. Users can view daily or weekly activity summaries, as well as a daily record of their locations and trips.

Viaggia Roveretgoto [15] provides gamification mechanisms to incentivize sustainable mobility choices. It integrates a journey planner that highlights in green the most sustainable options and presents them first. Users are rewarded with points based on the modes they use (including Green points for sustainable transportation, Health points for biking or walking and Park&Ride points for repeated park and ride facilities use).

From Table 2 we observe that self-monitoring is the most frequently used persuasion strategy and takes the form of visual feedback. The information being visualized is commonly the CO2 emissions caused by the users' trips ([11], [10], [12], [7], [9], [8]). Certain approaches provide visualizations of the cost and burned calories ([12], [14], [9]) calculated from the users' mobility patterns. The assumption is that when one switches to more environmentally friendly transport modes (e.g. from car to public transport or bicycle) the cost of mobility is reduced and users burn more calories.

Two forms of feedback are commonly used. Visual designs aim to communicate in a simple and user friendly manner aggregate statistics and take the form of cognitive representations of concepts that transform based on the users' activities. These con-cepts commonly trees that grow as users adopt more sustainable habits ([11], [8]), while in [8] an additional concept of a growing iceberg was used. The second form of feedback refers to charts (including bar and pie charts) presenting detailed statistics of the users' behaviour ([11], [14], [9], [12]).

Visual feedback is typically combined with and supports other persuasive strategies e.g. support for goal-setting and challenges), social comparison, inclusion of gamification and rewards (playful aspects), or personalized suggestions/messages.

Social aspects take the form of comparing the individual user performance to that of her/his peers (commonly other users who participate in the studies). The comparison can be initiated by the system which means that the visual feedback provides information on the user's performance compared to others, or social recognition with leaderboards that rank users according to their performance ([13], [15]). Additionally, it can be initiated by the user with functionalities to post one's performance in networks [3]. Rewards are given to users in the form of points for following sustainable modes ([11], [13], [15]). Persuasive messages are text based, whereas in [12] an approach that combines text with images is proposed.

## 4 Pilot Studies

Table 3 presents the main implementation details of the pilot studies examined, including the main goal, the number of users involved, the place where the study took place, and the way that goal achievement was evaluated. Not surprisingly, all studies have a similar goal: to promote sustainable mobility and to change travel behaviour using different persuasion strategies.

With respect to the number of users, it is observed that in most studies it is extremely small. In all cases, the number of participants is under 55, with the exception of studies [16] and [9]. In [16], although 471 participants used the prototype, a high dropout rate was observed, as only 65 of them completed both pre- and during-usage travel diaries. Moreover, in [16] users from three cities (i.e. Barcelona, Helsinki and Milan) participated, in contrast to the rest of the studies where participants were recruited from a single place. In [9], where 135 subjects participated in the pilot study, they were recruited from a subject pool of over 2500 UC Berkeley affiliates, most of which undergraduate students, while they also received a participation fee of $15 per hour.

The geographical spread of studies is rather small. Four of them took place in Italy, three in the USA, two in Austria and Finland, while one study was conducted in Spain. As far as the evaluation type is concerned, half of the studies ([16], [12], [7], [9], [15]) were evaluated only quantitatively through log analysis and/or questionnaires, while the rest of them ([11], [13], [10], [14] and [8]) were evaluated both quantitatively and qualitatively through interviews. In another classification, most of the studies were evaluated both objectively through log analysis and subjectively through questionnaires and/or interviews, with the exception of [15] which was evaluated only objectively and the studies [7] and [8] that were evaluated subjectively only.

**Table 3.** Main implementation details of the studies we examined including their focus, number of users, country and type of evaluation.

| Ref. | Details |
|---|---|
| [11] | *Focus:* To explore the effects of persuasive strategies and choice architecture in journey planning systems for sustainable transportation.<br>*Users:* 24, *Place:* Vienna – Austria. *Duration:* 8 Weeks. *Evaluation Type:* Analysis of system logs, Interviews. |
| [13] | *Focus:* To explore the impact of a journey planner app integrating goal-setting, self-monitoring, rewards and sharing features on transport choices and habits.<br>*Users:* 8, *Place:* Trento – Italy. *Duration:* 4 Weeks. *Evaluation Type:* Analysis of mobility habits logs, Interviews. |
| [10] | *Focus:* To explore users' reaction to actionable mobility challenges presented through a journey planning app.<br>*Users:* 12, *Place:* Helsinki – Finland. *Duration:* 4 Weeks. *Evaluation Type:* Analysis of app logs, Questionnaires before, during, and after the study, Interviews. |
| [16] | *Focus:* To assess the impact of a journey planner integrating goal-setting, self-monitoring and rewards.<br>*Users:* 471, *Place:* Barcelona – Spain, Helsinki - Finland, Milan – Italy. *Duration:* 3 weeks, *Evaluation Type:* Attitude and Motivation questionnaire, User experience questionnaire, Logs of travel diary app. |
| [12] | *Focus:* To understand the impact of persuasive information and advices delivered through mobile devices on car usage reduction.<br>*Users:* 15, *Place:* Cagliari – Italy. *Duration:* 2 weeks, *Evaluation Type:* Self-reporting Questionnaires, Logs of travel diary app. |
| [14] | *Focus:* To promote active modes such as bicycling and walking with the use of activity tracking smartphone applications.<br>*Users:* 35, *Place:* British Columbia, Minnesota - USA. *Duration:* 3 weeks, *Evaluation Type:* Activity logs, Interviews. |
| [7] | *Focus:* Persuade people to shift from using their cars to public transportation.<br>*Users:* 54, *Place:* Austria. *Duration:* N/A, *Evaluation Type:* Paratype method: prototype shown to participants in real life situations and immediately followed by a survey. |
| [9] | *Focus:* To explore whether travel feedback program can be replicated by a computational system.<br>*Users:* 135, *Place:* San Francisco Bay - USA. *Duration:* 3 weeks, *Evaluation Type:* Survey that measured factors that contribute to behaviour change, Analysis of trip logs. |
| [8] | *Focus:* To understand how participants react on visual feedback of CO2 emissions.<br>*Users:* 14, *Place:* Seattle - USA. *Duration:* 3 weeks, *Evaluation Type:* Online survey, experience sampling study. |
| [15] | *Focus:* To explore the potential of gamification mechanisms to incentivize behavioural changes towards sustainable mobility solutions.<br>*Users:* 40, *Place:* Rovereto - Italy. *Duration:* 5 weeks, *Evaluation Type:* Analysis of system logs containing the user selected trips. |

Table 4 presents the results reported in the studies we examined. Regarding the identified effectiveness of the approaches, the quantitative comparison of effect sizes is difficult due to huge differences in the methodological approach, sample size, geographic location and transportation context. It is interesting that none of the studies provides evidence of actual behavioural changes which can be attributed to the short durations of the pilots (limited to a maximum of a few weeks). However, the studies identify positive changes in users' perception of sustainability in urban mobility and increased concern regarding the impact of their choices on the environment.

Last but not least, based on the reported main findings and recommendations for future system designs of the analysed studies the following trends can be identified:

- Personalization is seen as an important possibility to improve the impact of the systems and to increase acceptability and real life usage. In the context of mobile persuasion aspects to address and personalize are especially route suggestions and alternatives.
- Localization of interventions, i.e. provide the right information in the right location is important.
- Good timing of interventions is important both for increasing the impact of the system and the acceptance by the users.
- Wearable devices might be a good starting point for unobtrusive in-context triggers for persuasive attempts.

**Table 4**. Overview of the results reported in the pilot studies.

| Ref. | Results |
| --- | --- |
| [11] | More positive attitudes towards environmentally friendly transport modes. User-reported behavioural changes, which could not be measured through logs analyses. |
| [13] | 14% improvement in the use of sustainable means of transport. Suggestion for longer periods of behaviour change intervention are required in order to replace transport habits. |
| [10] | Challenges were reported to affect consciousness of the consequences of user actions. Users pointed a personalization of the challenges would be more appropriate. |
| [16] | Social influence elements and personalized interventions are needed in order to increase the effectiveness of motivational features for large audiences. |
| [12] | Users indicated travel time as the most important feedback, followed by travel costs, $CO_2$ emitted and, lastly, calories burned. Persuasive messages containing text and cartoon images were preferred by participants. |
| [14] | No significant effects were observed with respect to increase of active modes usage. Users appreciated the always-on logging functionality of the provided system. |
| [7] | Significant correlations between opportune moment and possible attitude change and behaviour change which means that the right moment of system usage has a strong impact on the persuasive effect. |
| [9] | The effect of peer influences on attitudes enhances persuasion attempts. |
| [8] | Visual feedback increases awareness and consideration of the effects of travel on the environment. |
| [15] | Gamification can lead to the selection of sustainable routes. |

## 5    Discussion and Research challenges

In this section, we discuss the results obtained from the state-of-the-art analysis as well as the necessity and the impact of personalization, context awareness and proactivity in persuasion.

The results of the examined studies show that people differ in their susceptibility to different persuasive strategies. This leads to the assumption that personalized approaches can be more successful than "one size fits all" approaches. Many persuasive applications for sustainability have been implemented for a general audience using a single persuasive technique. For instance, IPET [12] motivates users to more eco-friendly habits providing visual feedback and sending personalized notifications. Thus, it is necessary to create services that address the needs of individual users (e.g. tailoring notifications). Personalization can also sustain users' interest over time by considering the different personality types. Some first results are encouraging, e.g. Jylhä et al., [10] reached better results by personalizing persuasive challenges, however further exploration of personalized persuasive strategies for behavioural change towards sustainable modes of transportation is required.

Another interesting observation concerns context awareness. Many persuasive applications can be considered context-aware, since they take contextual information into account while persuading users. However, they consider only one or two types of context such as location while ignoring other contextual data such as user context. Similarly, user's personalities which are types of context are ignored. Persuasive applications employ several persuasion techniques to motivate people promote more sustainable mobility. However, a large amount of research performed in the area of persuasive technologies widely acknowledges that some of these persuasion methods effect reversely some users. For example, competition as a kind of persuasion technique cannot motivate a broad range of people and that lose their appeal after a short period of time. By taking into account the personality of users, persuasive applications can tailor persuasion methods and therefore achieve more success.

Last, as indicated in [7], timely and proactive delivery of information can enhance the persuasive potential of an approach. Much of conventional choice theory assumes that each individual has complete knowledge of the alternatives and can make a rational choice. More recent empirical research [17] suggests that a much more proactive approach is required to not only inform individuals about the alternatives that are available but also help them decide which is most suitable for them. Information has to be provided to the user at the appropriate time, rather than assuming that they will find it themselves.

## 6    Conclusions

In this paper we reviewed persuasive systems and pilot studies related to behavioural change interventions for sustainable urban mobility. From our understanding personalized persuasion techniques extended with aspects of context awareness and proactivity should be realized in this field to move even the more reluctant urban traveller to shift from opportunistic choices to more environmentally aware and adopt sustainable travel behaviour. Currently, we are working in the EU funded project OPTIMUM (http://www.optimumproject.eu/) towards these directions. We are focusing on establishing persuasion profiles which will be used to tailor interventions to individual user characteristics. Moreover, we aim to take into account contextual information, including the personality and mood of the users and we are designing an approach to generate proactive notifications in order to deliver timely information and support user decisions when considering sustainable transportation modes.

## 7      Acknowledgements

Research reported in this paper has been partially funded by the European Commission project OPTIMUM (H2020 grant agreement no. 636160-2).